\documentclass[twocolumn]{pasj01}
\Received{}%{yyyy/mm/dd}
\Accepted{}%{yyyy/mm/dd}
\begin{document} 

\title{ 
%\LETTERLABEL %%% <-- uncomment for LETTER article  
%\REVIEWLABEL %%% <-- uncomment for REVIEW article  
Processing of hydrocarbon dust in star-forming galaxies revealed with AKARI}

%%% begin:list of authors
% Do NOT capitalize all letters in "textsc".
\author{Tsubasa \textsc{Kondo}\altaffilmark{1,}\footnotemark{}}
\email{t.kondo@u.phys.nagoya-u.ac.jp}

\author{Akino \textsc{Kondo}\altaffilmark{1}}
%\email{a.kondo@u.phys.nagoya-u.ac.jp}
\author{Katsuhiro L. M\textsc{urata}\altaffilmark{2}}

\author{Takuma \textsc{Kokusho}\altaffilmark{1}}
\author{Shinki \textsc{Oyabu}\altaffilmark{3}}
\author{Toyoaki \textsc{Suzuki}\altaffilmark{4}}
\author{Risako \textsc{Katayama}\altaffilmark{1}}
\author{Hidehiro \textsc{Kaneda}\altaffilmark{1}}

\altaffiltext{1}{Graduate School of Science, Nagoya University, Furo-cho, Chikusa-ku, Nagoya, Aichi, 464-8602, Japan}
\altaffiltext{2}{Okayama Observatory, Kyoto University, 3037-5 Honjo, Kamogata-cho, Asakuchi, Okayama, 719-0232, Japan }
\altaffiltext{3}{Institute of Liberal Arts and Sciences, Tokushima University, 1-1 Minami-josanjima-cho, Tokushima-shi, Tokushima, 770-8502, Japan}
\altaffiltext{4}{Institute of Space and Astronautical Science, Japan Aerospace Exploration Agency, 3-1-1 Yoshinodai, Chuo-ku, Sagamihara, Kanagawa, 252-5210 Japan}
%%% end:list of authors

%% `\KeyWords{}' always has to be placed before ``\maketitle'' 
%%  List of Key Words:  https://academic.oup.com/pasj/pages/Pasj_Keywords 

\maketitle

\begin{abstract}
%Please read ``IMPORTANT NOTICE'' carefully before preparing a manuscript. 
Hydrocarbon dust is one of the dominant components of interstellar dust, which mainly consists of polycyclic aromatic hydrocarbons and aliphatic hydrocarbons. While hydrocarbon dust is thought to be processed in interstellar radiation fields or shocks, detailed processing mechanisms are not completely understood yet. We investigate the processing of hydrocarbon dust by analyzing the relation between the luminosities emitted by hydrocarbon dust and the total infrared luminosities $(L_{\mathrm{IR}})$ for 138 star-forming galaxies at redshift $z\ <\ 0.3$. Using near-infrared 2.5--5~$\mathrm{\mu m}$ spectra obtained with AKARI, we derived the luminosities of the aromatic hydrocarbon feature at 3.3~$\mathrm{\mu m}$ ($L_\mathrm{aromatic}$) and the aliphatic hydrocarbon feature at 3.4--3.6~$\mathrm{\mu m}$ ($L_\mathrm{aliphatic}$). We also derived $L_\mathrm{IR}$ and the radiation field strength by modeling the spectral energy distributions of the 138 galaxies with AKARI, WISE and IRAS photometry data. We find that galaxies with higher $L_\mathrm{IR}$ tend to exhibit lower $L_\mathrm{aliphatic}/L_\mathrm{aromatic}$ ratios. Furthermore, we find that there is an anti-correlation between $L_\mathrm{aliphatic}/L_\mathrm{aromatic}$ ratios and the radiation field strength, and also that the galaxies with low $L_\mathrm{aliphatic}/L_\mathrm{aromatic}$ ratios are dominated by merger galaxies. These results support that hydrocarbon dust is processed through photodissociation in strong radiation fields and/or shocks during merging processes of galaxies; the $L_\mathrm{aliphatic}/L_\mathrm{aromatic}$ ratio is likely to decrease in such harsh interstellar conditions since the aliphatic bonds are known to be chemically weaker than the aromatic bonds.
\end{abstract}

%\pagewiselinenumbers

\section{Introduction}
%\noindent IMPORTANT NOTICE\\
%1. ``\verb|\draft|'' creates single column and double spaces format.\\
%2. If you comment out ``\verb|\draft|'', the output will be double column
%   and single space.\\
%3. For cross-references, the use of ``\verb|\label|, \verb|\ref|, \verb|\cite|'' 
%   and the thebibliography environment is strongly recommended. \\
%4. Do NOT use ``\verb|\def|, \verb|\renewcommand|''.\\
%5. Do NOT redefine commands provided by PASJ01.cls.\\

There are various types of interstellar dust with different sizes, compositions and structures. Hydrocarbon is one of the most important ingredients of interstellar dust, the presence of which is often identified through near- and mid-infrared (IR) spectral features due to vibrations of C--H or C--C bonds of aromatic or aliphatic hydrocarbons. Their smallest forms are known as polycyclic aromatic hydrocarbons (PAHs) or mixed aromatic-aliphatic organic nanoparticles (MAONs; \cite{k&z2011}, \yearcite{k&z2013}). It has also been proposed that small hydrocarbon dust is mostly hydrogenated amorphous carbon materials (a-C:H; \cite{jones2017}), which are associated with aliphatic to aromatic transformation. Both aromatic and aliphatic hydrocarbons are mostly composed of carbon and hydrogen, and hydrocarbon dust is thought to contain 50--1000 carbon atoms (\cite{tilens2008}). The primary difference between them is their structure; aromatic hydrocarbons contain benzene rings, while aliphatic hydrocarbons are organic chemical compounds which do not contain benzene rings. These hydrocarbons are excited by UV and optical radiation from stars and emit IR radiation in the near- to mid-IR wavelengths. The hydrocarbon emission is observed in a variety of objects and environments, including planetary nebulae, reflection nebulae, young stellar objects, star-forming galaxies and even quiescent galaxies (e.g., \cite{peeters2002}; \cite{brandl2006}; \cite{elbaz2011}; \cite{kaneda2005}; \cite{kaneda2008}). Aromatic hydrocarbons show strong emission features at 3.3, 6.2, 7.7, 8.6, 11.3 and 12.7~$\mathrm{\mu m}$ (e.g., \cite{li2020}), while aliphatic hydrocarbons show emission features at 3.4 and 6.9~$\mathrm{\mu m}$ (e.g., \cite{kwok2007}). The hydrocarbon features show wide diversities in the spectral position, shape and relative intensity. These diversities come from different properties of hydrocarbon dust, such as the chemical composition, ionization state and size distribution (e.g., \cite{d&l2007}; \cite{smith2007}; \cite{galliano2008}). \\
\indent Several past studies have reported the relationship between the hydrocarbon features and the interstellar conditions, and proposed the processing mechanism of hydrocarbon dust (e.g., \cite{galliano2008}, \cite{hemahcandra2015}). For example, for the nearby edge-on starburst galaxy M82, \citet{yamagishi2012} indicated that the aliphatic to aromatic flux ratio increases with the distance from the galactic center. They concluded that shocks by galactic superwinds produce small carbonaceous grains which emit the aliphatic features through fragmentation of larger carbonaceous grains, while dense molecular clouds protect aromatic hydrocarbons from the fragmentation near the galactic center region. \citet{pilleri2015} found that the aliphatic to aromatic intensity ratio decreases with the intensity of the radiation field in the reflection nebula NGC 7023, indicating that aliphatic C--H bonds are photodissociated by strong UV radiation. For hydrocarbon dust around young stars, \citet{acke2010} also found that the aliphatic C--H bonds are more easily destroyed than the aromatic C--H bonds by UV photons. In Galactic H\,\emissiontype{II} regions, \citet{mori2014} found that the ratio of the aromatic to the aliphatic hydrocarbon intensity decreases with the ionization degree of PAH that is affected by the UV radiation field, which is likely to be caused by the fact that aliphatic hydrocarbons are more fragile than aromatic hydrocarbons. As a result of the JWST (James Webb Space Telescope; \cite{gardner2023}) observation of the nearby galaxy NGC 7469, \citet{lai2023} found that the ratio of the aromatic to the aliphatic intensity varies spatially in its star-forming ring. Toward the galactic nucleus of NGC 7469, the ratio shows a decreasing trend, which again indicates that the aliphatic hydrocarbons are preferentially destroyed by harder radiation fields around its active galactic nucleus (AGN). Hence it has been shown that the aliphatic to the aromatic flux ratios are significantly variable, indicating that hydrocarbon dust can be processed in interstellar radiation fields and/or shocks. Yet, there are few statistical studies on the relationships between the aliphatic and aromatic components for a large number of objects with different interstellar conditions. \\
\indent Among past studies closely related to the present one, \citet{yamada2013} confirmed that the aromatic hydrocarbon to the total infrared luminosity ratios decrease with the total infrared luminosities for 101 star-forming galaxies at redshift $z\ <\ 0.3$. This trend was particularly seen in the ultra-luminous infrared galaxies (ULIRGs: $L_\mathrm{IR}\ >\ 10^{12}\ L_{\odot}$). They suggested that shocks induced by galaxy mergers might have destroyed aromatic hydrocarbons. Moreover, \citet{murata2017} presented evidence that aromatic hydrocarbons are destroyed not only by shocks but also by strong radiation fields, analyzing the physical properties of merger and non-merger galaxies. In this paper, we systematically explore the relation of the aliphatic and aromatic hydrocarbon dust features with physical properties of star-forming galaxies, using a large number of near-IR spectra (2.5--5~$\mathrm{\mu m})$ obtained with the IRC (Infrared Camera; \cite{onaka2007}) aboard AKARI (\cite{murakami2007}).

\section{Data analysis}
\subsection{AKARI mid-IR excess sample}
Our sample consists of 230 galaxies, which are based on the AKARI mission programs, Mid-infrared Search for Active Galactic Nuclei (MSAGN: \cite{oyabu2011}) and Evolution of ULIRGs and AGNs (AGNUL: \cite{imanishi2008}, \yearcite{imanishi2010}). These two programs performed near-IR spectroscopy at 2.5--5~$\mathrm{\mu m}$ wavelengths with a spectral resolution of $R \sim 120$ using AKARI/IRC (\cite{ohyama2007}). In \citet{yamada2013}, 184 galaxies were selected according to the criterion of $F(9\ \mathrm{or}\ 18\ \mathrm{\mu m})$/$F(K_\mathrm{s})\ >\ 2$, where $F(9\ \mathrm{or}\ 18\ \mathrm{\mu m})$ and $F\left(K_\mathrm{s}\right)$ are the flux densities at 9 or 18~$\mathrm{\mu m}$ and in the $K_\mathrm{s}$ band obtained with AKARI and 2MASS, respectively. In this study, we re-evaluated $F(9\ \mathrm{or}\ 18\ \mathrm{\mu m})$ with photometry using the latest version of the AKARI mid-IR all-sky map (\cite{ishihara2010}). As a result, the number of sample galaxies (230) has increased by 46 from the sample in \citet{yamada2013}. These sample galaxies are mid-IR-excess sources, which are regarded as infrared galaxies with active star formation and/or AGNs.

\subsection{Near-IR spectral fitting}
We carry out model fitting to the AKARI/IRC 2.5--5~$\mathrm{\mu m}$ spectra which include various spectral features of the interstellar medium, such as aromatic hydrocarbon, aliphatic hydrocarbon, $\mathrm{H_2O}$ ice, and atomic and molecular hydrogen. In order to study the hydrocarbon features, we model the spectra as follows:
\begin{eqnarray}
\label{full_screen_model}
F_{\nu}\ =\ \mathrm{e}^{-\tau}\left(F_\mathrm{continuum}\ +\ F_\mathrm{aromatic}\ +\ F_\mathrm{aliphatic}\right),
\end{eqnarray}
where $\tau$ is the optical depth for the $\mathrm{H_2O}$ ice absorption, and $F_\mathrm{continuum},\ F_\mathrm{aromatic}$ and $F_\mathrm{aliphatic}$ are the flux densities of the near-IR continuum, aromatic hydrocarbon feature at 3.3~$\mathrm{\mu m}$ and aliphatic hydrocarbon features at 3.4--3.6~$\mathrm{\mu m}$, respectively. We assume the full screen geometry for the $\mathrm{H_2O}$ ice absorption. For the near-IR continuum, we use a power-law model, which represents thermal dust and stellar continua and is defined as $F_\mathrm{continuum} \propto \lambda^{\Gamma}$. We use a Drude profile (\cite{l&d}) for the 3.3~$\mathrm{\mu m}$ aromatic hydrocarbon feature, which is defined as
\begin{eqnarray}
F_\mathrm{aromatic}\ =\ \frac{a b^{2}}{\left[(\lambda / \lambda_{\mathrm{r}})\ -\ (\lambda_{\mathrm{r}} / \lambda\right)]^{2}\ +\ b^{2}},
\end{eqnarray}
where $\lambda_{\mathrm{r}}$, $a$ and $b$ are the center wavelength of the feature, the peak flux density at $\lambda_{\mathrm{r}}$ and the full width at half maximum (FWHM) divided by $\lambda_{\mathrm{r}}$, respectively. We fix $\lambda_{\mathrm{r}}$ at 3.3~$\mathrm{\mu m}$ in the rest frame. The aliphatic hydrocarbon features at 3.4--3.6~$\mathrm{\mu m}$ are thought to be composed of several components. Here we assume 4 representative components, the center wavelengths of which are 3.41, 3.46, 3.51 and 3.56~$\mathrm{\mu m}$. We use a Lorentzian profile for the 3.41~$\mathrm{\mu m}$ feature and Gaussian profiles for the 3.46, 3.51 and 3.56~$\mathrm{\mu m}$ features (\cite{mori2012}; \cite{kondo2012}). These models are selected depending on whether or not the emission features are spectrally resolved with the AKARI/IRC. We fix the widths of the 3.46, 3.51 and 3.56~$\mathrm{\mu m}$ features at the instrumental resolution of 0.034~$\mathrm{\mu m}$. Finally, we use a Lorentzian profile for the $\mathrm{H_2O}$ ice absorption feature, the center wavelength of which is fixed at 3.05~$\mathrm{\mu m}$. With this fitting, we calculate the aromatic and aliphatic hydrocarbon luminosities ($L_\mathrm{aromatic}$ and $L_\mathrm{aliphatic}$).

\subsection{Spectral energy distribution fitting}
We carry out the fitting of spectral energy distributions (SEDs), using near- to far-IR photometric data obtained with AKARI, WISE (Wide-field Infrared Survey Explorer; \cite{wright2010}) and IRAS (Infrared Astronomical Satellite; \cite{neugebauer1984}). We use the same photometric data and SED model based on DustEM (\cite{compiegne2011}) as presented in \citet{murata2017}. The SED model consists of the following three dust components: PAHs, amorphous silicate (aSil) and hydrogenated amorphous carbon (amC). PAHs are composed of neutral and ionized PAHs. We fit the SEDs of the sample galaxies on the condition that the abundance ratio of the neutral to ionized PAHs is free. The amC is divided into large amC (LamC) and small amC (SamC) based on the dust size. In addition, we include two blackbody components with the temperatures of 600~K and 3000~K, which represent a hot dust and a stellar continuum, respectively. Finally, we calculate the total infrared luminosity ($L_\mathrm{IR}$) and the radiation field strength ($G_{0}$) to estimate the physical environments of the sample galaxies. Here, $G_{0}$ is the far-UV (6.0--13.6~eV) radiation field strength normalized to the solar neighborhood value (\cite{habing1968}).

\subsection{The final sample}
\label{final_sample}
\begin{figure*}[h]
\begin{center}
\includegraphics[bb=0 0 553 136, width=15.8cm]{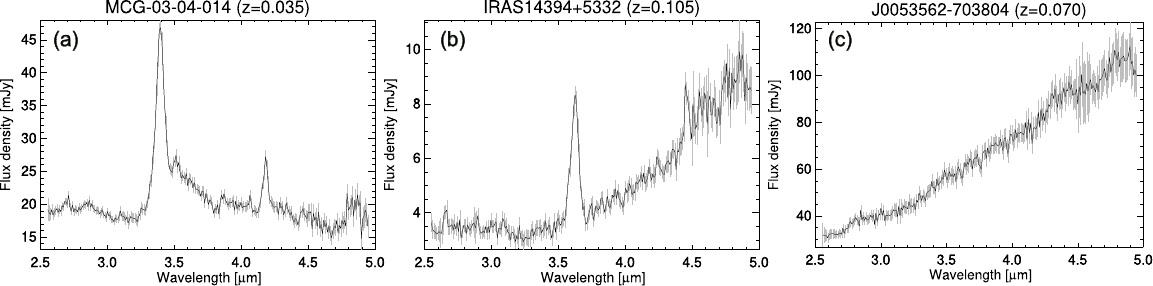}
\end{center}
\caption{Examples of the AKARI/IRC 2.5-5~$\mathrm{\mu m}$ spectra: (a) star-forming galaxy, (b) composite galaxy and (c) AGN-dominated galaxy.}
\label{spectrum_exam}
\end{figure*}

\begin{figure*}[h]
\begin{center}
\includegraphics[bb=0 0 583 139, width=15.8cm]{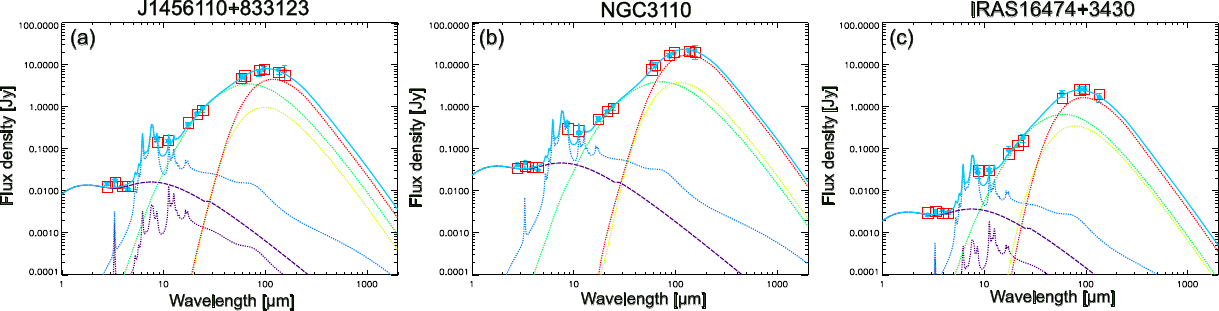}
\end{center}
\caption{Examples of the SED fitting results for star-forming galaxies: (a) IRG, (b) LIRG and (c) ULIRG. The observed flux densities are indicated with filled cyan circles. The SED model is composed of ionized PAHs (blue dotted line), neutral PAHs (purple dotted line), SamC (green dotted line), LamC (yellow dotted line), aSil (red dotted line) and near-IR continuum (purple dashed line). The total model spectrum and the model flux densities are shown with a solid cyan line and open red squares, respectively.}
\label{SED_fitting_exam}
\end{figure*}

We classify the sample galaxies into three categories according to the star formation activity and the presence of AGNs evaluated by the near-IR spectral fitting. If star formation is active in a galaxy, we detect a strong 3.3~$\mathrm{\mu m}$ aromatic hydrocarbon feature. On the other hand, if there is an AGN in a galaxy, PAHs are destroyed in strong radiation fields near the AGN, and therefore the 3.3~$\mathrm{\mu m}$ aromatic hydrocarbon feature is expected to be suppressed. Hence, we regard the equivalent width of the 3.3~$\mathrm{\mu m}$ aromatic hydrocarbon feature, $\mathrm{EW_{3.3\ \mu m}}$, as a tracer of the star formation activity. In this study, the star formation is considered to be active in the case of $\mathrm{EW_{3.3\ \mu m}}\ >\ 40\ \mathrm{nm}$ (\cite{moorwood1986}; \cite{imanishi2000}; \cite{imanishi2008}). Moreover, AGNs make near-IR continuum slopes steeper due to the contribution of a hot dust continuum. Therefore, we regard the near-IR continuum slope, $\Gamma$ ($F_\mathrm{continuum}\ \propto\ \lambda^{\Gamma}$), as a standard of whether or not the contamination of AGNs is strong. In this study, the contamination of AGNs is considered to be strong in the case of $\Gamma>1$ (\cite{risaliti2006}; \cite{imanishi2010}). Consequently, the sample galaxies are classified into the following 3 categories: (1) star-forming galaxies ($\mathrm{EW_{3.3\ \mu m}}\ >\ 40\ \mathrm{nm}$ and $\Gamma\ <\ 1$), (2) composite galaxies which show both star-forming and AGN activities ($\mathrm{EW_{3.3\ \mu m}}\ >\ 40\ \mathrm{nm}$ and $\Gamma\ >\ 1$) and (3) AGN-dominated galaxies ($\mathrm{EW_{3.3\ \mu m}}\ <\ 40\ \mathrm{nm}$ and $\Gamma\ >\ 1$). Figure \ref{spectrum_exam} shows examples of the spectra thus categorized. \\
\indent Furthermore, we also classify the sample galaxies into another 3 categories according to their $L_\mathrm{IR}$ derived from the SED fitting. The galaxies are classified into the following 3 categories: (1) infrared galaxies (IRGs; $L_\mathrm{IR}\ <\ 10^{11}\ L_\odot$), (2) luminous infrared galaxies (LIRGs; $10^{11}\ L_\odot\ <\ L_\mathrm{IR}\ <\ 10^{12}\ L_\odot$ ) and (3)  ultra-luminous infrared galaxies (ULIRGs; $L_\mathrm{IR}\ >\ 10^{12}\ L_\odot$). Figure \ref{SED_fitting_exam} shows examples of the SED fitting results for star-forming galaxies belonging to these categories. \\
\indent Finally, our sample contains 138 star-forming galaxies, 21 composite galaxies and 19 AGN-dominated galaxies. We excluded the other 52 galaxies from our original sample of the 230 galaxies because they were not fitted well by the spectral model or the SED model on a $\chi^2$ test with a 90~\% confidence level. The spectral types and the luminosity classes are summarized in table \ref{galaxies_classification}. In this paper, we discuss the 138 star-forming galaxies, which consist of 15 IRGs, 74 LIRGs and 49 ULIRGs.

\begin{table*}[h]
\tbl{Classification of the sample galaxies.}{%
\begin{tabular}{ccccc}
\hline
Criteria & Star-forming & Composite & AGN & Total \\
\hline
Star formation activity ($\mathrm{EW_{3.3\ \mu m}}\ >\ 40\ \mathrm{nm}$) & Yes & Yes & No & - \\
\hline
Near-IR continuum slope ($\Gamma\ >\ 1$) & No & Yes & Yes & - \\
\hline
%\multicolumn{5}{c}{Sample} \\
Sample & \multicolumn{4}{l}{} \\
\hline
IRGs ($L_\mathrm{IR}\ <\ 10^{11}\ L_\odot$) & 15 & 1 & 3 & 19 \\
\hline
LIRGs ($10^{11}\ L_\odot\ <\ L_\mathrm{IR}\ <\ 10^{12}\ L_\odot$) & 74 & 5 & 8 & 87 \\
\hline
ULIRGs ($L_\mathrm{IR}\ >\ 10^{12}\ L_\odot$) & 49 & 15 & 8 & 72 \\
\hline
Total & 138 & 21 & 19 & 178 \\
\hline
\end{tabular}}
\label{galaxies_classification}
\end{table*}

\section{Results}
\begin{figure*}[h]
\begin{tabular}{cc}
\begin{minipage}[c]{.45\linewidth}
\centering
\includegraphics[bb=0 0 461 300, width=9.0cm]{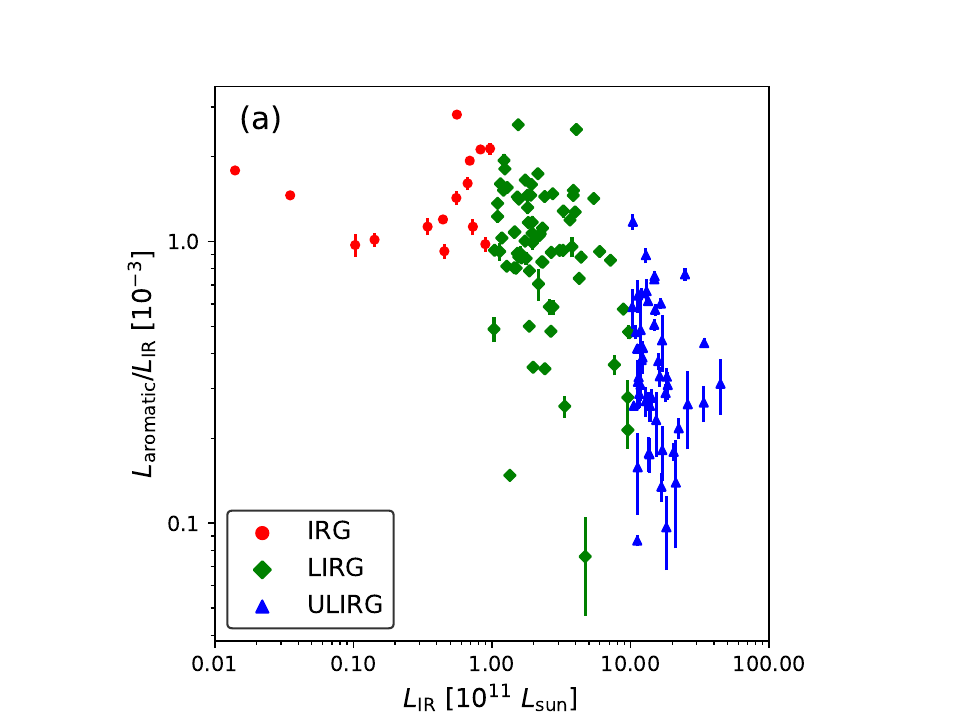}
\end{minipage}
\begin{minipage}[c]{.45\linewidth}
\centering
\includegraphics[bb=0 0 461 300, width=9.0cm]{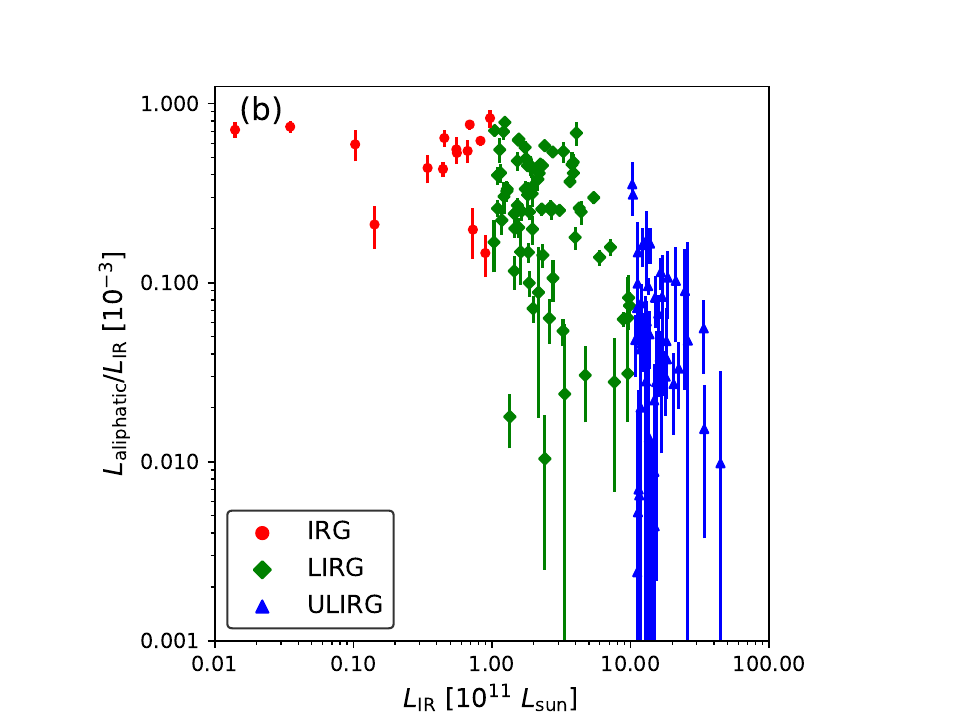}
\end{minipage}
\end{tabular}
\caption{(a) $L_\mathrm{aromatic}/L_\mathrm{IR}$ and (b) $L_\mathrm{aliphatic}/L_\mathrm{IR}$ plotted against $L_\mathrm{IR}$ for the star-forming galaxies: IRGs (red circles), LIRGs (green diamonds) and ULIRGs (blue triangles).}
\label{laro_lali_lir}
\end{figure*}

We examine the relations between the hydrocarbon emission luminosities ($L_\mathrm{aromatic}$ and $L_\mathrm{aliphatic}$) and the total infrared luminosities ($L_\mathrm{IR}$) for the 138 star-forming galaxies. Figures \ref{laro_lali_lir}a and \ref{laro_lali_lir}b display $L_\mathrm{aromatic}/L_\mathrm{IR}$ and $L_\mathrm{aliphatic}/L_\mathrm{IR}$, respectively, plotted against $L_\mathrm{IR}$. As already reported in \citet{yamada2013}, $L_\mathrm{aromatic}/L_\mathrm{IR}$ significantly decreases with $L_\mathrm{IR}$ at high $L_\mathrm{IR}$ ($\geq\ 10^{11}\ L_\odot$), while $L_\mathrm{aromatic}/L_\mathrm{IR}$ shows a nearly constant value ($L_\mathrm{aromatic}/L_\mathrm{IR}\ \sim\ 10^{-3}$) at low $L_\mathrm{IR}$ ($<\ 10^{11}\ L_\odot$). As a new result, we find that $L_\mathrm{aliphatic}/L_\mathrm{IR}$ also significantly decreases with $L_\mathrm{IR}$, similarly to the trend of $L_\mathrm{aromatic}/L_\mathrm{IR}$. These results are reasonable for the interpretation that both aromatic and aliphatic hydrocarbons are of the same carriers for the band emission (\cite{k&z2011}, \yearcite{k&z2013}).

\begin{figure}[h]
\begin{center}
\includegraphics[bb=50 0 400 320, width=7.5cm]{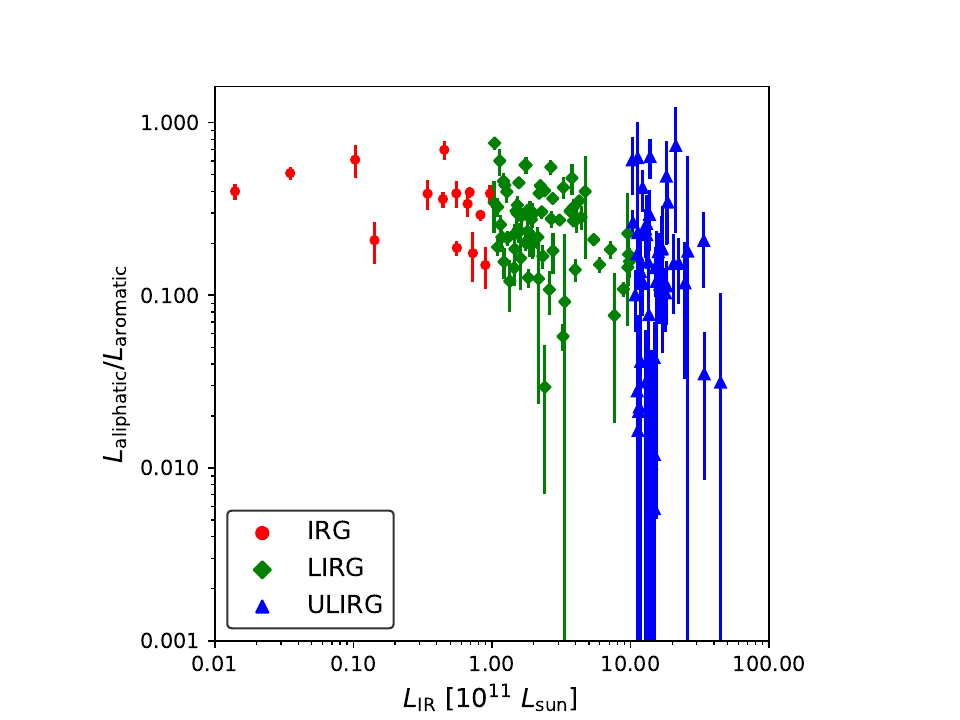}
\end{center}
\caption{$L_\mathrm{aliphatic}/L_\mathrm{aromatic}$ plotted against $L_\mathrm{IR}$ for the star-forming galaxies: IRGs (red circles), LIRGs (green diamonds) and ULIRGs (blue triangles).}
\label{aliaro_lir}
\end{figure}

\begin{figure*}[h]
\begin{center}
\includegraphics[bb=0 0 568 162, width=15.8cm]{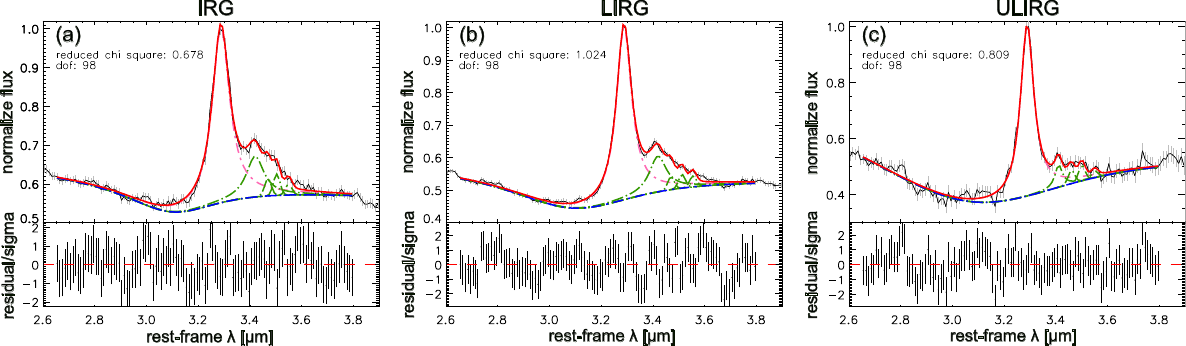}
\end{center}
\caption{AKARI/IRC 2.6-3.9~$\mathrm{\mu m}$ stacked spectra (black solid line with error bars) of the star-forming galaxies: (a) IRGs, (b) LIRGs and (c) ULIRGs. The spectral model is composed of the near-IR continuum (blue dot-dashed line), aromatic hydrocarbon feature (pink dot-dashed line) and aliphatic hydrocarbon features (green lines), all of which are influenced by the $\mathrm{H_2O}$ ice absorption. The total model spectrum is shown with a red solid line.}
\label{stack_spectral_pasj}
\end{figure*}

Figure \ref{aliaro_lir} shows the relationship between $L_\mathrm{aliphatic}/L_\mathrm{aromatic}$ and $L_\mathrm{IR}$. As can be seen in the figure, $L_\mathrm{aliphatic}/L_\mathrm{aromatic}$ shows a decreasing trend with a large scatter at higher $L_\mathrm{IR}$. For a robustness check of this trend, we carry out model fitting for the three spectra which are created by stacking the spectra at the wavelengths of 2.6--3.9~$\mathrm{\mu m}$ in their rest frames for each luminosity class, normalizing the flux densities at 3.3~$\mathrm{\mu m}$, the peak wavelength of the aromatic hydrocarbon feature. Figure \ref{stack_spectral_pasj} displays the result of fitting the spectra stacked for IRGs, LIRGs and ULIRGs, where the best-fit $L_\mathrm{aliphatic}/L_\mathrm{aromatic}$ are 0.262 $\pm$ 0.012, 0.247 $\pm$ 0.007 and 0.111 $\pm$ 0.012 for IRGs, LIRGs and ULIRGs, respectively. As shown in figure \ref{stack_spectral_pasj}, $L_\mathrm{aliphatic}/L_\mathrm{aromatic}$ indeed decreases with the luminosity class, and thus it is likely that hydrocarbon dust tends to be more aromatic-rich in galaxies with higher $L_\mathrm{IR}$.

\section{Discussion}
Below we discuss main causes to generate the trend of $L_\mathrm{aliphatic}/L_\mathrm{aromatic}$ as seen in figure \ref{aliaro_lir}.

\subsection{The effect of the absorption feature due to carbonaceous dust}
\begin{figure}[h]
\begin{center}
\includegraphics[bb=50 0 400 330, width=7.5cm]{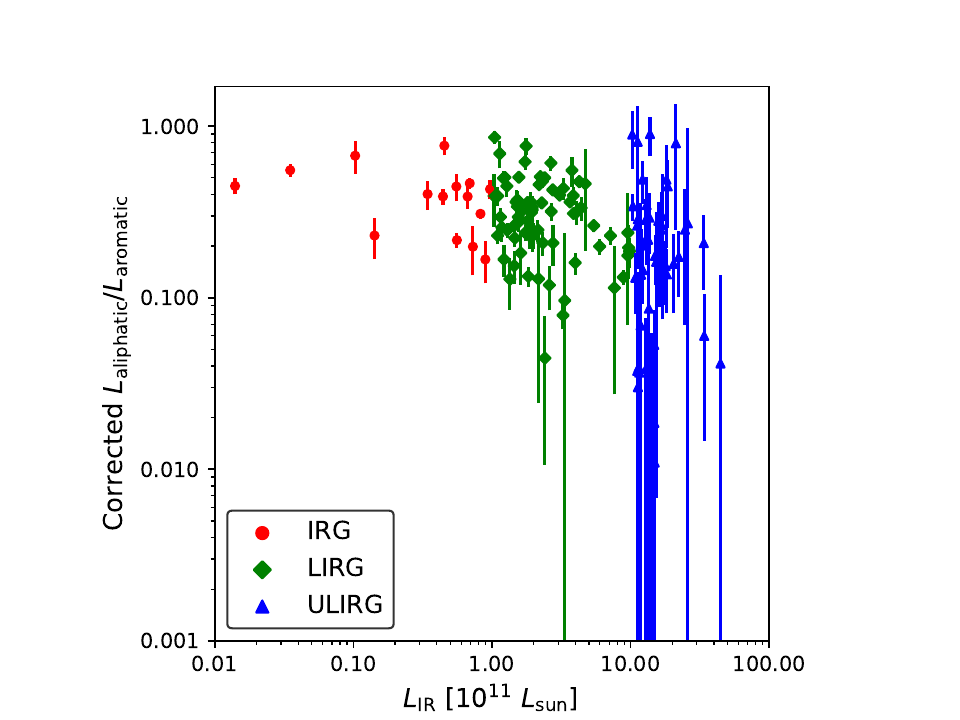}
\end{center}
\caption{Same as figure \ref{aliaro_lir}, but absorption-corrected $L_\mathrm{aliphatic}/L_\mathrm{aromatic}$ derived from full screen geometry model plotted against $L_\mathrm{IR}$ for the star-forming galaxies: IRGs (red circles), LIRGs (green diamonds) and ULIRGs (blue triangles).}
\label{abs_aliaro_lir}
\end{figure}

First of all, in order to investigate whether or not the variations of $L_\mathrm{aliphatic}/L_\mathrm{aromatic}$ are really intrinsic, we evaluate the effect of the absorption feature due to carbonaceous dust. For example, a significant fraction of dusty AGNs are known to exhibit the absorption feature at the wavelength of 3.4~$\mathrm{\mu m}$ due to cooler carbonaceous dust (e.g., \cite{imanishi2008}, \yearcite{imanishi2010}). If the absorption is significant in our sample, the tendency between $L_\mathrm{aliphatic}/L_\mathrm{aromatic}$ and $L_\mathrm{IR}$ could be affected, since we do not consider the 3.4~$\mathrm{\mu m}$ absorption feature in our spectral fitting. It is difficult to separately estimate the aliphatic hydrocarbon emission and the carbonaceous dust absorption in the spectral fitting. Instead, we assume that the optical depth of carbonaceous dust, $\tau_\mathrm{3.4\ \mu m}$, correlates with the optical depth of $\mathrm{H_2O}$ ice, $\tau_\mathrm{3.1\ \mu m}$, where we take $\tau_\mathrm{3.4\ \mu m}/\tau_\mathrm{3.1\ \mu m}$ of 0.85 as derived from AKARI observations (\cite{imanishi2008}, \yearcite{imanishi2010}), and estimate $\tau_\mathrm{3.4\ \mu m}$ from the measured $\tau_\mathrm{3.1\ \mu m}$. Figure \ref{abs_aliaro_lir} shows the relationship between $L_\mathrm{aliphatic}/L_\mathrm{aromatic}$ thus corrected for the possible presence of the absorption feature and $L_\mathrm{IR}$, from which we find that the absorption-corrected $L_\mathrm{aliphatic}/L_\mathrm{aromatic}$ still decreases with $L_\mathrm{IR}$ significantly. Hence this result rules out a possibility that the carbonaceous dust absorption is the main cause of the decrease in $L_\mathrm{aliphatic}/L_\mathrm{aromatic}$, verifying that $L_\mathrm{aliphatic}/L_\mathrm{aromatic}$ is intrinsically variable.

\subsection{The effect of photodissociation in strong radiation fields}
\begin{figure}[h]
\begin{center}
\includegraphics[bb=50 0 400 330, width=7.5cm]{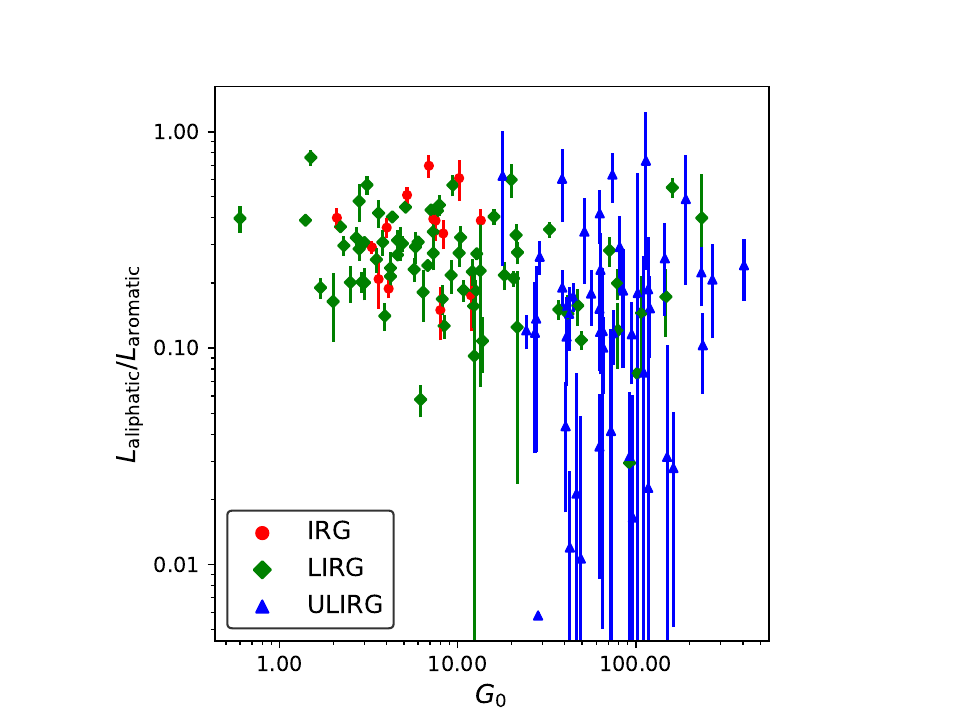}
\end{center}
\caption{$L_\mathrm{aliphatic}/L_\mathrm{aromatic}$ plotted against $G_0$ for the star-forming galaxies: IRGs (red circles), LIRGs (green diamonds) and ULIRGs (blue triangles).}
\label{aliaro_g0}
\end{figure}

Here and hereafter we consider the processing mechanism of hydrocarbon dust to change $L_\mathrm{aliphatic}/L_\mathrm{aromatic}$. First, we evaluate the effect of strong radiation fields using $G_0$ derived from the SED fitting. Figure \ref{aliaro_g0} shows $L_\mathrm{aliphatic}/L_\mathrm{aromatic}$ plotted against $G_0$. All the galaxies with very low $L_\mathrm{aliphatic}/L_\mathrm{aromatic}$ ($<\ 0.05$) show $G_0$ larger than 30, indicating that there is some threshold on $G_0$ to significantly reduce $L_\mathrm{aliphatic}/L_\mathrm{aromatic}$. The threshold on $G_0$ suggests that the strong radiation field is likely to cause the decrease in $L_\mathrm{aliphatic}/L_\mathrm{aromatic}$, since the aliphatic bonds are known to be chemically weaker than the aromatic bonds for photodissociation. Yet there is a large scatter in $L_\mathrm{aliphatic}/L_\mathrm{aromatic}$ at $G_0$ larger than 30 where some galaxies show relatively high $L_\mathrm{aliphatic}/L_\mathrm{aromatic}$. Hence we cannot explain the variation of $L_\mathrm{aliphatic}/L_\mathrm{aromatic}$ by $G_0$ alone.

\subsection{The effect of shocks with galaxy mergers}
\begin{figure*}[h]
\begin{tabular}{cc}
\begin{minipage}[b]{.45\hsize}
\centering
\includegraphics[bb=0 0 461 320, width=9.0cm]{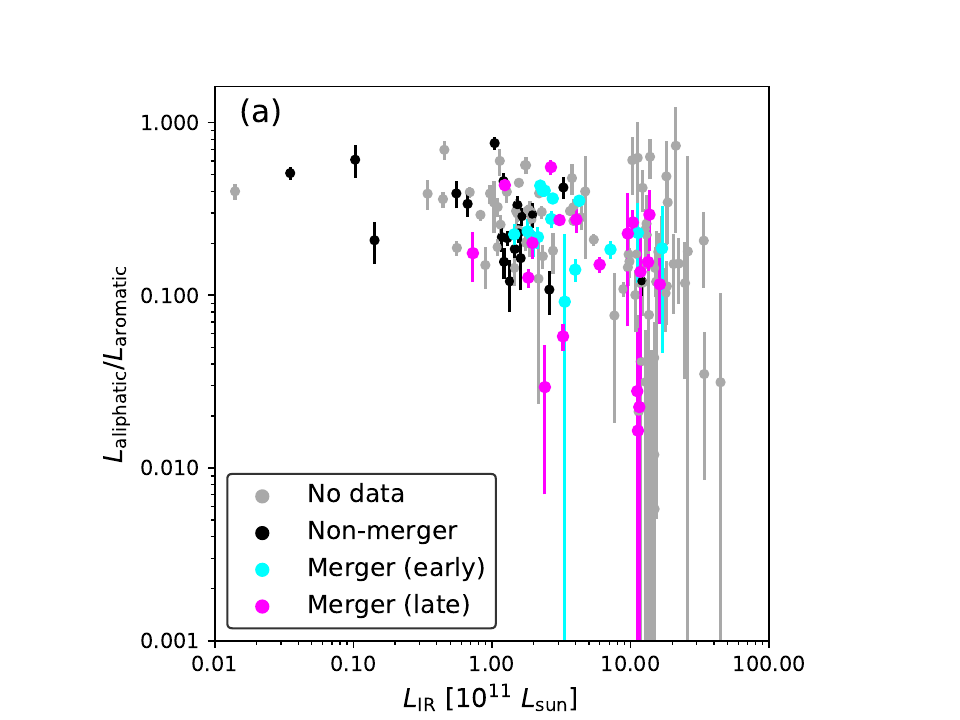}
\end{minipage}
\begin{minipage}[b]{.45\hsize}
\centering
\includegraphics[bb=0 0 461 320, width=9.0cm]{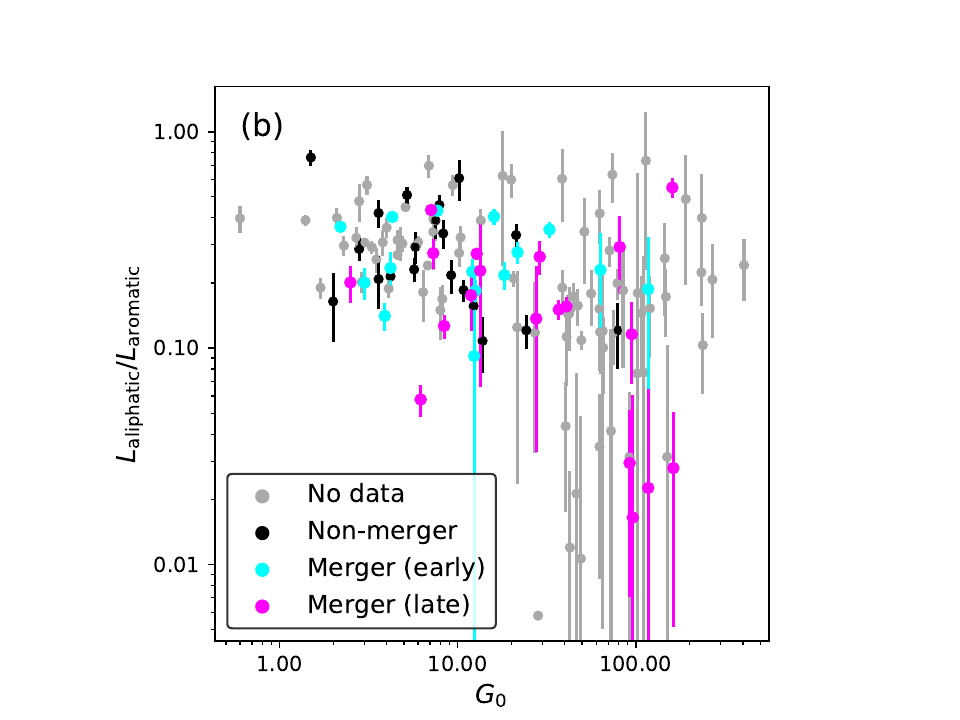}
\end{minipage}
\end{tabular}
\caption{(a) Same as figure \ref{aliaro_lir} and (b) same as figure \ref{aliaro_g0}, but color-coded for non-merger galaxies (black circles), early-stage merger galaxies (light blue circles) and late-stage merger galaxies (magenta circles), as classified in \citet{murata2017}.}
\label{aliaro_lir_merger}
\end{figure*}

Next, we explore the relationship between $L_\mathrm{aliphatic}/L_\mathrm{aromatic}$ and the galaxy merger. Most galaxies with high $L_\mathrm{IR}$ like ULIRGs are thought to have experienced galaxy mergers. The merger of gas-rich galaxies triggers starburst with large-scale shocks, considerably affecting the interstellar environments of the galaxies. It is generally known that the scale and the speed of the shocks increase at later stages of the galaxy merger (\cite{rich2015}). \citet{murata2017} investigated the presence and the stage of the galaxy merger for their sample of 55 star-forming galaxies, all of which are included in our galaxy sample and categorized as 15 early-stage merger galaxies, 20 late-stage merger galaxies and 20 non-merger galaxies. Figure \ref{aliaro_lir_merger}a shows the relationship between $L_\mathrm{aliphatic}/L_\mathrm{aromatic}$ and $L_\mathrm{IR}$ as classified with the merger stage. The average values of $L_\mathrm{aliphatic}/L_\mathrm{aromatic}$ for non-merger galaxies, early-stage merger galaxies and late-stage merger galaxies are $0.306\ \pm\ 0.175$, $0.263\ \pm\ 0.104$ and $0.176\ \pm\ 0.144$, respectively. According to a t-test with a 95\% confidence level, there is a significant difference in the distribution of $L_\mathrm{aliphatic}/L_\mathrm{aromatic}$ between the non-merger galaxies and the merger galaxies; the group of aliphatic hydrocarbons may selectively be removed in merger galaxies, causing a change in the structure of hydrocarbon dust. On the other hand, there is no significant difference in the distribution of $L_\mathrm{aliphatic}/L_\mathrm{aromatic}$ between the early-stage merger galaxies and the late-stage merger galaxies, according to a t-test with a 95\% confidence level. Therefore, although the galaxy merger is likely to be one of the main causes for the processing of hydrocarbon dust, its dependence on the merger stage is not clear.

\begin{figure}[h]
\begin{center}
\includegraphics[bb=45 0 400 330, width=7.5cm]{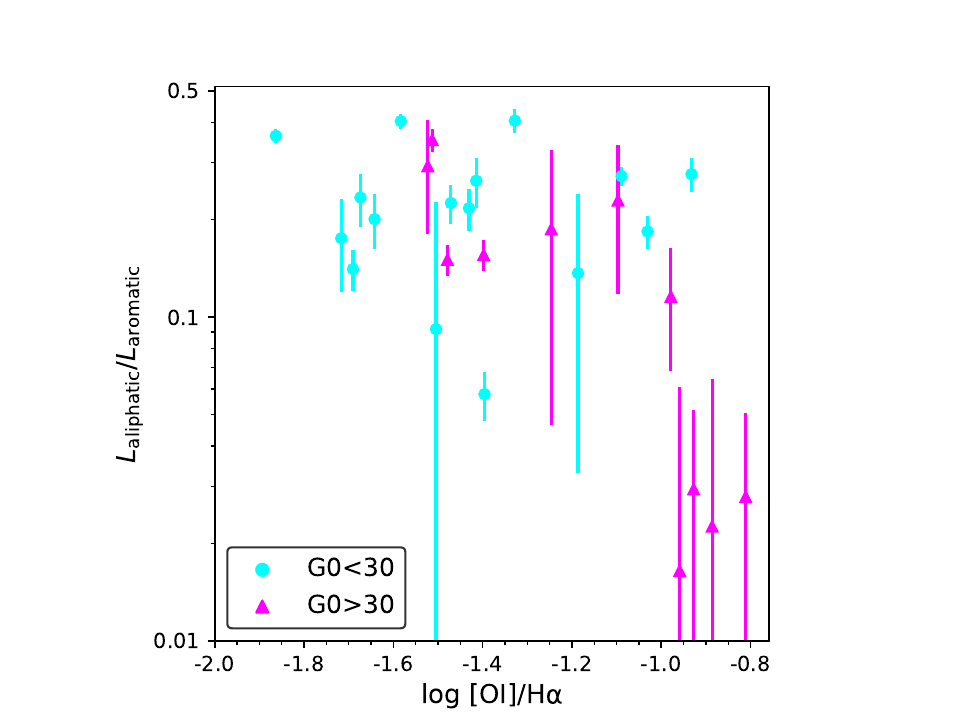}
\end{center}
\caption{$L_\mathrm{aliphatic}/L_\mathrm{aromatic}$ plotted against $\mathrm{[O\,\emissiontype{I}]/H\alpha}$ for the star-forming galaxies: merger galaxies with $G_\mathrm{0}$ below the threshold ($G_\mathrm{0}\ =\ 30$) (light blue circles) and merger galaxies with $G_\mathrm{0}$ above the threshold (magenta triangles)}
\label{aliaro_OIHa}
\end{figure}

Figure \ref{aliaro_lir_merger}b shows the relationship between $L_\mathrm{aliphatic}/L_\mathrm{aromatic}$ and $G_\mathrm{0}$ as classified with the merger stage. As can be seen in the figure, above the threshold of $G_\mathrm{0}\ =\ 30$, there is a clear difference in $L_\mathrm{aliphatic}/L_\mathrm{aromatic}$ between the early-stage merger galaxies and the late-stage merger galaxies, although the statistics are rather poor. Only late-stage merger galaxies show very low $L_\mathrm{aliphatic}/L_\mathrm{aromatic}$ ($<\ 0.05$) at high $G_\mathrm{0}$ ($>\ 30$), and thus it is likely that strong shocks significantly affect the structure of hydrocarbon dust at the late stage of the galaxy merger. \\
\indent In order to further probe the environments of the merger galaxies, we examine the relationship between $L_\mathrm{aliphatic}/L_\mathrm{aromatic}$ and $\mathrm{[O\,\emissiontype{I}]\lambda6300\AA/H\alpha}$, a shock tracer as used in \citet{murata2017}, in figure \ref{aliaro_OIHa}, where we find that $L_\mathrm{aliphatic}/L_\mathrm{aromatic}$ is anti-correlated with $\mathrm{[O\,\emissiontype{I}]/H\alpha}$ for the merger galaxies with high $G_\mathrm{0}$ ($>\ 30$). Therefore, the strength of the shock in the galaxy merger is indeed a likely cause to modify the structure of hydrocarbon dust.

\begin{figure*}[h]
\begin{tabular}{cc}
\begin{minipage}[b]{.45\hsize}
\centering
\includegraphics[bb=0 0 461 320, width=9.0cm]{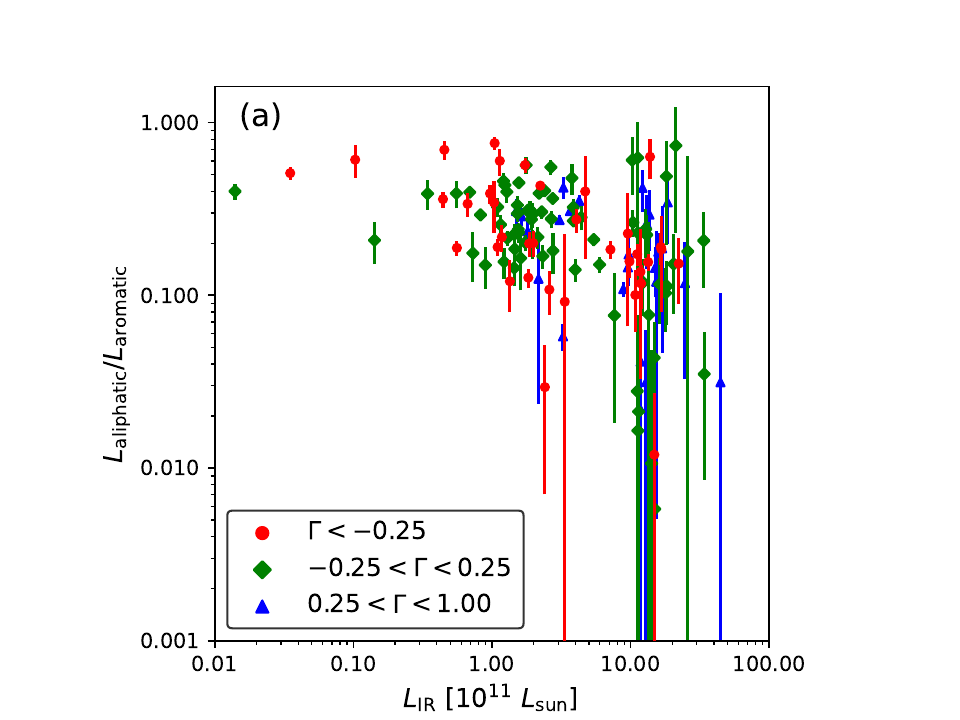} 
\end{minipage}
\begin{minipage}[b]{.45\hsize}
\centering
\includegraphics[bb=0 0 461 320, width=9.0cm]{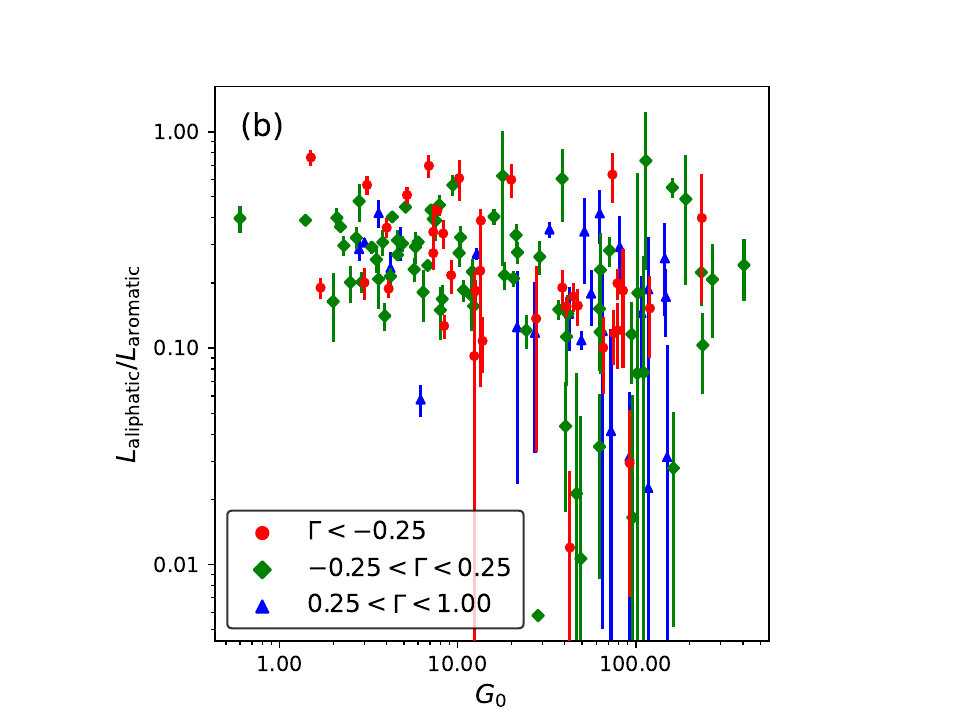} 
\end{minipage}
\end{tabular}
\caption{(a) Same as figure \ref{aliaro_lir} and (b) same as figure \ref{aliaro_g0}, but color-coded with $\Gamma$: $\Gamma\ <\ -0.25$ (red circles), $-0.25\ <\ \Gamma\ <\ 0.25$ (green diamonds) and $0.25\ <\ \Gamma\ <\ 1.00$ (blue triangles)}
\label{aliaro_lir_gamma_comp}
\end{figure*}

\begin{figure*}[h]
\begin{tabular}{cc}
\begin{minipage}[b]{.45\hsize}
\centering
\includegraphics[bb=0 0 461 330, width=9.0cm]{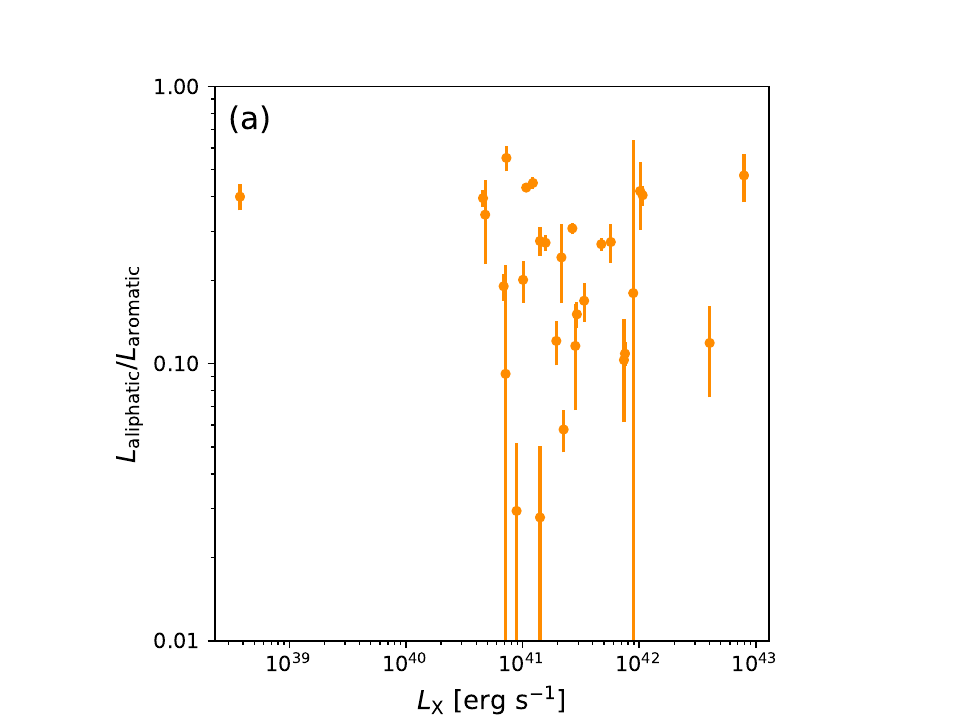} 
\end{minipage}
\begin{minipage}[b]{.45\hsize}
\centering
\includegraphics[bb=0 0 461 330, width=9.0cm]{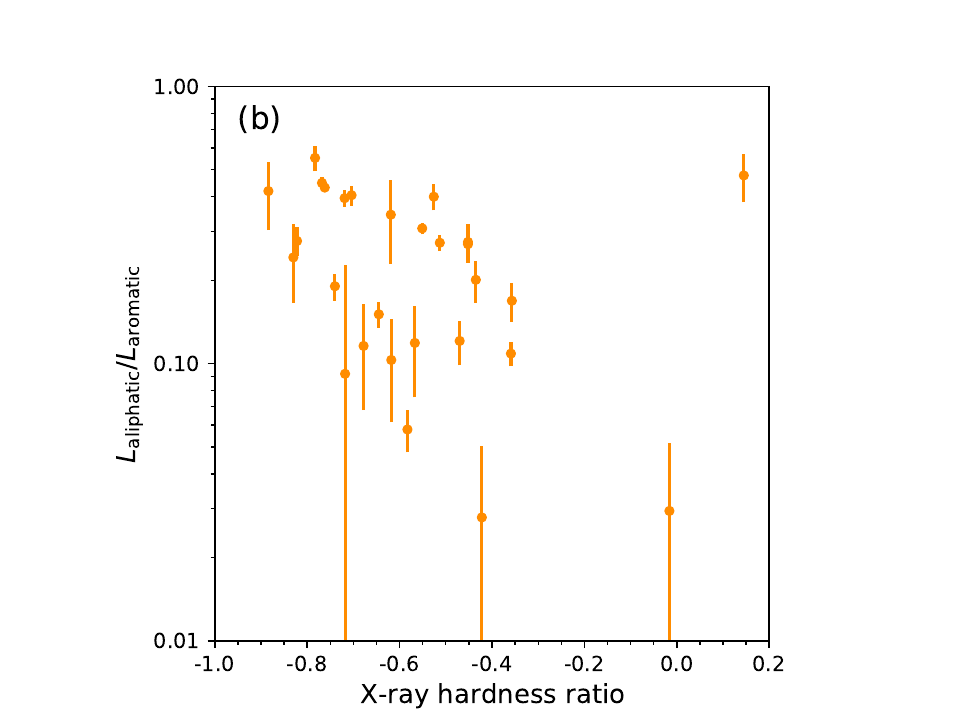}
\end{minipage}
\end{tabular}
\caption{$L_\mathrm{aliphatic}/L_\mathrm{aromatic}$ plotted against (a) $L_\mathrm{X}$ and (b) the X-ray hardness ratio for the star-forming galaxies.}
\label{aliaro_xray_comp}
\end{figure*}

\subsection{A possible contribution of AGNs in the star-forming galaxies}
Based on the near-IR continuum slope, $\Gamma$, which is expected to reflect the degree of the contribution from AGNs, we have removed AGNs (i.e., AGN-dominated and composite galaxies) from our sample. Nevertheless, the remaining star-forming galaxies, especially those with $\Gamma$ near the threshold ($\Gamma\ =\ 1$) for the classification, may have a non-negligible contribution of relatively weak AGNs, which can still process and destroy the hydrocarbon dust. In order to check this possibility, we investigate the dependence of $L_\mathrm{aliphatic}/L_\mathrm{aromatic}$ on $\Gamma$ to evaluate a possible contribution of such AGNs in the star-forming galaxies. \\
\indent In figures \ref{aliaro_lir_gamma_comp}a and \ref{aliaro_lir_gamma_comp}b, we show the same plots as figures \ref{aliaro_lir} and \ref{aliaro_g0}, respectively, but color-coded with $\Gamma$, in which we confirm that there is no systematic difference in the $L_\mathrm{aliphatic}/L_\mathrm{aromatic}$--$L_\mathrm{IR}$ and $L_\mathrm{aliphatic}/L_\mathrm{aromatic}$--$G_\mathrm{0}$ distributions with different $\Gamma$. Hence it is unlikely that our sample star-forming galaxies suffer a significant contribution of weak AGNs. \\
\indent Moreover, we match our sample with the XMM-Newton X-ray source catalog (\cite{rosen2016}). We find that 27 star-forming galaxies in our sample are detected in the X-ray. Using the information on the X-ray luminosity, $L_\mathrm{X}$, and the X-ray hardness ratio, HR, the latter of which is defined as 
\begin{eqnarray}
\mathrm{HR}\ =\ \frac{\mathrm{CR_{2.0-12.0\ keV}}\ -\ \mathrm{CR_{0.2-2.0\ keV}}}{\mathrm{CR_{2.0-12.0\ keV}}\ +\ \mathrm{CR_{0.2-2.0\ keV}}},
\end{eqnarray}
where $\mathrm{CR_{0.2-2.0\ keV}}$ and $\mathrm{CR_{2.0-12.0\ keV}}$ are the X-ray photon count rates of XMM-Newton/EPIC in the energy bands of 0.2--2.0~keV and 2.0--12.0~keV, respectively. Figure \ref{aliaro_xray_comp} shows the relationships between $L_\mathrm{aliphatic}/L_\mathrm{aromatic}$ and these X-ray parameters. In figure \ref{aliaro_xray_comp}a, we find that $L_\mathrm{aliphatic}/L_\mathrm{aromatic}$ is not correlated with $L_\mathrm{X}$, confirming that the decrease in $L_\mathrm{aliphatic}/L_\mathrm{aromatic}$ is not caused by AGNs which are expected to have high $L_X$. On the other hand, figure \ref{aliaro_xray_comp}b shows that $L_\mathrm{aliphatic}/L_\mathrm{aromatic}$ is anti-correlated with the X-ray hardness ratio. Besides AGNs, supernovae in the starburst caused by the galaxy merger can also increase the X-ray hardness ratio in the star-forming galaxies, where shocks and hot plasma environments associated with young supernova remnants may have processed the hydrocarbon dust to the aromatic-rich.

\section{Conclusion}
Based on the luminosities of the aromatic hydrocarbon feature at 3.3~$\mathrm{\mu m}$ ($L_\mathrm{aromatic}$) and the aliphatic hydrocarbon feature at 3.4--3.6~$\mathrm{\mu m}$ ($L_\mathrm{aliphatic}$) estimated with the AKARI 2.5--5~$\mathrm{\mu m}$ spectra, we have studied the relationships between $L_\mathrm{aromatic}$, $L_\mathrm{aliphatic}$ and the total infrared luminosity, $L_\mathrm{IR}$, for 138 star-forming galaxies. As a result, we find that galaxies with higher $L_\mathrm{IR}$ (i.e., LIRGs and ULIRGs) tend to exhibit lower $L_\mathrm{aliphatic}/L_\mathrm{aromatic}$ ratios. The galaxies with low $L_\mathrm{aliphatic}/L_\mathrm{aromatic}$ ratios are dominated by merger galaxies, showing strong radiation fields ($G_0$) estimated with the total infrared SED and strong shocks traced by $\mathrm{[O\,\emissiontype{I}]/H\alpha}$. We search for any non-negligible effect of AGNs contaminating the star-forming galaxy sample to find no clear signature for that, except for significant anti-correlation of $L_\mathrm{aliphatic}/L_\mathrm{aromatic}$ against the X-ray hardness ratio. The overall results indicate that hydrocarbon dust is destroyed through photodissociation in strong radiation fields and/or shocks during the merging processes of galaxies. Since the aliphatic bonds are known to be chemically weaker than the aromatic bonds, the group of aliphatic hydrocarbons may selectively be removed, causing a change in the structure of hydrocarbon dust to the aromatic-rich.

%%%%%%%%%%%%%%%%%%%%%%%%%%%%%%%%%%%%%%%
\begin{ack}
%We thank an anonymous referee for her/his careful reading of our manuscript and for giving useful comments. 
This research is based on observations with AKARI, a JAXA project with the participation of ESA. This publication makes use of data products from the Wide-field Infrared Survey Explorer, which is a joint project of the University of California, Los Angeles, and the Jet Propulsion Laboratory/California Institute of Technology, funded by the National Aeronautics and Space Administration, and data products from the Infrared Astronomical Satellite (IRAS), which is a joint project of the US, UK and the Netherlands. This work is financially supported by JST SPRING, Grant Number JPMJSP2125. The principal author thanks the “THERS Make New Standards Program for the Next Generation Researchers.”
\end{ack}

%%%
% See the manual for the detail.
%%%

\end{document}